# Phonon studies of the phase transition sequence in antiferroelectric single crystal of Pb(Hf$_{0.83}$Sn$_{0.17}$)O$_3$

Anirudh K.R.[1,2], Cosme Milesi-Brault[1,3], Christelle Kadlec[1], Dmitry Nuzhnyy[1], Andrzej Majchrowski[4], Magdalena Krupska-Klimczak[5,6], Irena Jankowska-Sumara[6] and Elena Buixaderas[1*]

[1] Institute of Physics, Czech Academy of Sciences, Na Slovance 2, 182 00 Prague 8, Czech Republic

[2] Faculty of Nuclear Sciences and Physical Engineering, Czech Technical University, 11519, Prague, Czech Republic

[3] Université Paris-Saclay, CentraleSupélec, Laboratoire SPMS, 91190 Gif-sur-Yvette, France

[4] Institute of Applied Physics, Military University of Technology, ul. Kaliskiego 2, 00-908 Warsaw, Poland.

[5] Institute of Security Studies and Computer Science, University of the National Education Commission, Podchorążych 2, 30-084 Kraków, Poland

[6] Faculty of Exact and Natural Sciences, University of the National Education Commission, Podchorążych 2, 30-084 Kraków, Poland

* buixader@fzu.cz

**Abstract**

The sequence of phase transitions in PbHf$_{0.83}$Sn$_{0.17}$O$_3$ has been studied by THz, far infrared and Raman spectroscopies, revealing the complementary behaviour of both, polar and non-polar phonons and their impact on the transition lattice dynamics. Pb atom is sensitive to all phase transitions, changing its dynamics with temperature. As temperature decreases, the crystal undergoes a sequence of three phase transitions. The first one to an intermediate (IM) phase, in which polar fluctuations are detected by THz and IR spectroscopy at frequencies below 100 cm$^{-1}$ contributing to the maximum of permittivity and revealing important softening. Additional antipolar Pb fluctuations and softening were detected by Raman spectroscopy. At lower temperature, another transition to an antiferroelectric (AFE2) phase is revealed by nonpolar soft modes (antipolar and antiferrodistortive ones), and by an important drop of the dielectric strength of polar phonons. The final transition to the antiferroelectric phase (AFE1) is revealed by the appearance of new modes and a sudden change in the frequency of the soft modes when the antipolar shifts become larger.

Using symmetry analysis and optical observation to study how the domain pattern changes with temperature, we identified a path for the cubic-AFE2 transition throughout the IM phase, of plausible tetragonal symmetry, driven by an instability from the center of the Brillouin zone. This mechanism coexists with antiferrodistortive instabilities that eventually drive the material into the AFE2 phase. The final phase transition to the AFE1 phase naturally follows from a mode outside the center of the Brillouin zone and a further doubling of the unit cell.

# 1. Introduction

Antiferroelectric materials are gradually becoming a primary focus in materials science research due to their distinctive properties and potential applications in various fields. The study of antiferroelectricity and the understanding of its phase transition mechanism are nowadays very active research areas in both fundamental and applied sciences as they can play a pivotal role in advancing energy storage, sensing, cooling, and optoelectronic technologies [1], [2], [3].

The most popular antiferroelectric perovskite oxide is $PbZrO_3$. Lattice dynamic investigations on this material have uncovered valuable information related to the nature of the antiferroelectric phase transitions [4], [5]. The multiple lattice instabilities of $PbZrO_3$ make this material prone to developing other phases at high temperatures depending on the atoms introduced, and specifically on the site these atoms are placed. For instance, titanium in the B-site forming the solid solution $Pb(Zr_{1-x}Ti_x)O_3$ (PZT), leads to ferroelectricity [6], and lanthanum substitution in the A-site forming other solid solution $(Pb_{1-x}La_x)(Zr_{1-y}Ti_y)O_3$ (PLZT) leads to incommensurate modulations and to the strengthening of antiferroelectricity [7].

Recently, $PbHfO_3$, structurally analogous to $PbZrO_3$, has gained attention for its potential in energy storage applications [8], [9], [10], an area that has been explored less than in $PbZrO_3$, but in which $PbHfO_3$ can outperform $PbZrO_3$ in terms of energy storage capability, possibly due to the smaller size of Hf ions [8]. The main advantage of using Hf instead of Zr is the decrease of polar nature in the structure and the strengthening of the antiferroelectric character. Hf introduces a middle phase between the cubic paraelectric and the low temperature antiferroelectric orthorhombic phase. As a result, $PbHfO_3$, exhibits two successive antiferroelectric (AFE) phase transitions below the cubic paraelectric (PE) phase: one at ~470 K to a middle phase (AFE2), and another at ~430 K to the orthorhombic phase (AFE1) with *Pbam* space group [11], [12], [13]. The AFE2 phase, seen below ~470 K, is antiferroelectric and orthorhombic. However, due to the intrinsic disorder introduced by Hf, this phase is incommensurate with the *Imma*(00γ)s00 super-space group [14]. This incommensurability is linked to the displacement of lead ions and the tilting of the $HfO_6$ octahedra. The modulation vector depends slightly on temperature, affecting the lattice dynamics of the material.

By introducing Sn into the Hf site of $PbHfO_3$, forming $PbHf_{1-x}Sn_xO_3$ (PHS-X, with X=100x), the energy storage performance can be significantly improved [15] and new phases have been discovered [16] similar to those observed in $PbZrO_3$ substituted with Sn [17]. Introducing $Sn^{4+}$ ions into $PbHfO_3$ has a similar effect to applying hydrostatic pressure. This isovalent substitution can be considered as a form of chemical pressure since the ionic radii of $Sn^{4+}$ ions are slightly smaller than those of the ions they replace. Thus, Sn atoms in PHS solid solutions help stabilizing the intermediate AFE2 phase which occurs naturally in PbHfO$_3$ [11-14,18]. For compounds with $x \geqslant 0.08$, Sn substitution leads to the formation of an additional intermediate (IM) phase between the PE and the high temperature antiferroelectric phase AFE2, whose origin and symmetry are still under debate [19]. This IM phase resembles the high-pressure β phase reported in [18]. As with hydrostatic pressure, Sn$^{4+}$ ions promote instability in the oxygen sublattice related to octahedral tilting, which seems to play a crucial role in driving phase transitions. The full transition sequence from the cubic phase is PE (cubic)-IM-AFE2 (orthorhombic *Imma* and IC)-AFE1 (orthorhombic *Pbam*), as shown in the phase diagram [20]. The new IM phase is characterized by fluctuations of oxygen octahedra tilts and Pb disorder, probably in the form of antipolar (or antiparallel) uncorrelated shifts, where atoms in neighboring cells shift the same amount in opposite directions [19], [21]. Brillouin spectroscopy indicates that the average symmetry of the intermediate phase is cubic, composed of fine ferroelastic domains [22]. Raman experiments have shown that phonons in PHS with various compositions are highly sensitive to the phase transitions, exhibiting strong phonon softening in the AFE2 phase [20], [23]. However, no infrared experiments demonstrating the presence of polar phonons in these antiferroelectric crystals have been reported so far.

As the exploration of PHS and similar complex perovskites is fundamental for advancing the field of antiferroelectric perovskites to obtain materials with enhanced antiferroelectricity and energy-storage properties, this paper aims to deepen the understanding of changes induced by Sn substitution and its impact on the lattice dynamics of PbHfO$_3$. For this purpose, we use IR, THz, and Raman spectroscopies on a PHS crystal with $x$=0.17. We focus on comparing polar and non-polar vibrational modes, and their impact on the lattice dynamics across phase transitions. With the aid of symmetry analysis and temperature-dependent observations of the domain evolution, we establish a possible pathway for the PE-IM-AFE2 transitions and propose a plausible symmetry for the IM phase.

## 2.     Materials and experimental methods

Single crystals of PbHf$_{0.83}$Sn$_{0.17}$O$_3$ were grown by a self-flux method using 2.4 mol% PbHf$_{1-x}$Sn$_x$O$_3$, 77 mol% Pb$_3$O$_4$, and 20.6 mol% B$_2$O$_3$ as starting materials. Initially, the Sn composition was set at 24 mol% substituting Hf. Ultimately, a solid solution with the chemical formula PbHf$_{0.83}$Sn$_{0.17}$O$_3$ was obtained, determined by energy-dispersive X-ray spectroscopy, and it is referred to as PHS-17 in the text for convenience. A more detailed description of the crystallization process can be found in ref [24]. The resulting crystals were transparent plates, slightly greyish in color, with dimensions of several mm$^2$ and thickness less than 1 mm.

For specific heat measurements, a single piece of crystal with a mass of ~30 mg was placed in an aluminium crucible. The measurements were performed using a Netzsch DSC F3 Maia calorimeter, over a temperature range of 120–550 K. Temperature ramps were conducted at a rate of 1 K/min. The values of transition energies $\Delta E$ and excess entropy $\Delta S$ were calculated by the integration of the excess heat capacity $\Delta c_p(T)$ as:

$$\Delta E = \int \Delta c_p(T) dT, \quad \Delta S = \int \frac{\Delta c_p(T)}{T} dT$$

where $c_p$ is the heat capacity at constant pressure.

Electric permittivity measurements were performed using an Agilent 4363 LCR meter and a programmable temperature controller (Lake Shore, model 331). Measured crystals in the form of thin plates were coated with silver electrodes and placed in a furnace, where the temperature was controlled by a thermocouple with an accuracy of 0.1 K. The temperature rate was set to 1 K/min.

Far infrared (IR) reflectivity spectra were acquired using a Fourier transform IR spectrometer Bruker IFS 113v equipped with pyroelectric detectors as well as a He-cooled (1.6 K) Si bolometer. Room temperature spectra were measured in the range 30–3000 cm$^{-1}$ (~$10^{12}$–$10^{14}$ Hz). For low-temperature measurements (from 300 to 10 K), a continuous He-flow Oxford Optistat CF cryostat was used, with the sample mounted in a He-gas bath. Due to the polyethylene windows of the cryostat, the upper-frequency limit was restricted to ~600 cm$^{-1}$. To polarize the light, we utilized a metal-mesh polarizer deposited on a thin polyethylene foil. For high temperatures (300–700 K), an adapted commercial high-temperature cell Specac P/N 5850 was placed inside the spectrometer.

Time-domain terahertz transmission spectroscopy (TDTTS) from 0.2 to 2 THz (~6–60 cm$^{-1}$) was performed using a custom-made time-domain THz transmission spectrometer, equipped with an Optistat CF cryostat with Mylar windows for measurements down to 10 K, and an adapted commercial furnace Specac P/N 5850 for high temperatures up to 700 K. For the TDTTS experiment the crystal was glued onto a transparent sapphire substrate and then thinned down to 23 μm. This procedure allows to measure the transmitted THz field without breaking the sample. Because of the multilayer configuration, it is necessary to measure independently the properties of the substrate, and then extract the values of the index of refraction of the crystal by modeling the system as a multilayer structure. For details about the custom-made setup see [25] and, for the THz related calculations on thin films [26].

Raman scattering measurements were performed in a back-scattering geometry using a RM-1000 Renishaw Raman microscope equipped with a double Bragg filter, providing an excellent rejection of the laser beam below 5 cm$^{-1}$. The Raman spectra were collected using the 514.5 nm line of an Ar laser at a power of 25 mW (~4 mW on the sample) using a long-distance 20× objective. The diameter of the laser spot on the sample surface amounted to ~10 μm. The spectral resolution was better than 1.5 cm$^{-1}$ with a 2400 l/mm grating. To control the temperature of the samples a THMS-600 cell (LINKAM) was used during the cooling, from 600 K to room temperature. All Raman spectra were subsequently corrected for the Bose-Einstein thermal population factor.

Optical microscopy with polarized white light was used to observe the evolution of domain patterns in transmission during both heating and cooling cycles using the same temperature-controlled cell.

## 3. Results

*3.1 Differential scanning calorimetry (DSC) and dielectric measurements*

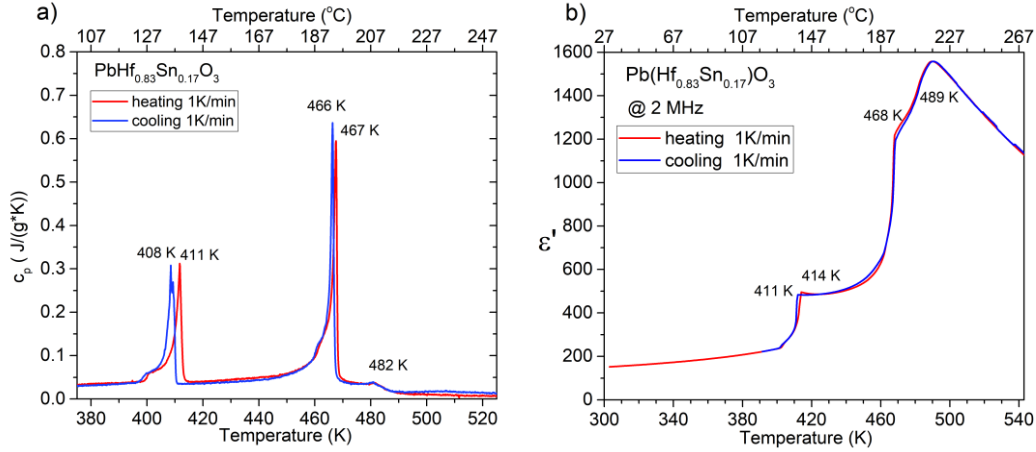

Figure 1: a) Temperature dependence of the heat capacity $c_p$ of the PHS-17 crystal on heating and cooling at the rate 1K/min. b) Temperature dependence of the permittivity $\varepsilon'$ of PHS-17 at 2 MHz on heating and cooling at the rate 1K/min

The results of the calorimetric measurements on heating and cooling (Figure 1a) evidence several phase transitions. The two strong $c_p$ peaks correspond to transitions between the two antiferroelectric phases (AFE1-AFE2, at ~410 K) and between the second antiferroelectric phase and the intermediate phase (AFE2-IM, at ~465 K), as the temperature increases A small peak at the highest temperatures, near 480 K, indicates the transition between the intermediate phase and the cubic paraelectric phase (IM-PE, at ~482 K). The sequence of the phase transitions is consistent with an earlier phase diagram [20]. The occurring of temperature hysteresis at the AFE1-AFE2 phase transition is characteristic for first-order phase transitions. The same applies to the phase transition between the AFE2 and IM phases. The smaller anomaly related to the IM-PE phase transition, which lacks thermal hysteresis, provides evidence of a second-order continuous phase transition between these two phases, as previously confirmed in PHS solid solutions by structural investigations [20,21]. A closer look reveals that the two prominent peaks have small shoulders on the left. Similar behavior has been previously observed for similar crystals of the PHS and PbZr$_{1-x}$Sn$_x$O$_3$ families, with various Sn concentrations [16, 27]. The presence of such shoulders on the $c_p(T)$ curves near phase transitions was linked with the existence of temperature ranges in which two phases coexist. This phase coexistence is a characteristic feature of first-order phase transitions when a new phase starts to develop within the current phase. Further evidence of how phase transitions occur in this crystal is provided later, where changes in the domain structure are described (see also Suppl. Material).

The associated changes in enthalpy ΔH and entropy ΔS for PHS-17 at the AFE1-AFE2 phase transition were estimated to be ΔH ~430 J/mol and ΔS ~1.15 J/(mol·K). The values at AFE2-IM are higher: ΔH ~860 J/mol and ΔS ~1.9 J/(mol·K). Both calculated ΔS values are much smaller than the theoretical value for an order-disorder phase transition —for a simple model with two sites, ΔS = Rln (N) ~ 5.8 J/(mol·K), using N=2 and the ideal gas constant R=8.314 J/mol·K) [28]—, which suggests that both phase transitions have an important displacive contribution.

Dielectric measurements at 2 MHz also reveal the same sequence of phase transitions (Figure 1b) with transition temperatures similar to those observed by DSC. As the crystal showed no dispersion within the measured frequency range (200 Hz–2 MHz), only permittivity values at 2 MHz are displayed.

According to the phase diagram [20], the increasing amount of Sn ions in $PbHfO_3$ leads to a decrease of temperatures of both AFE1-AFE2 and AFE2-IM phase transitions, whereas the temperature of the main phase transition at $T_C$ increases. The existence of the IM phase is manifested by an initially minor, then significant, flattering of the ε'(T) dependence a few degrees below the main phase transition at $T_C$. At the same time, the maximum value of the dielectric permittivity systematically decreases with the increasing amount of Sn ions.

*3.2    IR reflectivity and THz transmission measurements*

The IR reflectivity of the PHS-17 crystal was measured from 520 K to 10 K. The unpolarized spectra are displayed in Figure 2. Data from the THz experiment (TDTTS) were added after the proper calculation of the reflectivity from the measured complex permittivity. Spectra were normalized with respect to the points measured in the THz range, as this experiment is done in transmission and it is not so sensitive to the reference data, as the reflectivity experiment.

The corresponding fits are also shown in the figure, according to a model in which the reflectivity coefficient is calculated by

$$R(\omega) = \left|\frac{\sqrt{\hat{\varepsilon}(\omega)}-1}{\sqrt{\hat{\varepsilon}(\omega)}+1}\right|^2, \qquad (1)$$

where $\hat{\varepsilon}(\omega)$ is the complex dielectric function.

The spectra of complex perovskites are characterized by very broad and asymmetric bands, especially at high temperatures. Due to this the spectra were fitted using the generalized or 4-parameter oscillator model and the factorized form of the dielectric function [29]:

$$\hat{\varepsilon}(\omega) = \varepsilon'(\omega) - i\varepsilon''(\omega) = \varepsilon_\infty \prod_{i=1}^n \frac{\omega_{LOi}^2 - \omega^2 + i\omega\gamma_{LOi}}{\omega_{TOi}^2 - \omega^2 + i\omega\gamma_{TOi}} \qquad (2)$$

where $\varepsilon_\infty$ is the permittivity at frequencies much higher than the phonon frequencies, $\omega_{TOi}$ and $\omega_{LOi}$ are the transverse and longitudinal optical frequencies of the *i*-th phonon mode, and $\gamma_{TOi}$ and $\gamma_{LOi}$ their respective damping parameters. The contribution of each oscillator is calculated by

$$\Delta\varepsilon_i = \frac{\varepsilon_\infty}{\omega_{\text{TO}i}^2} \frac{\prod_j (\omega_{\text{LO}j}^2 - \omega_{\text{TO}i}^2)}{\prod_{j \neq i} (\omega_{\text{TO}j}^2 - \omega_{\text{TO}i}^2)} \qquad (3)$$

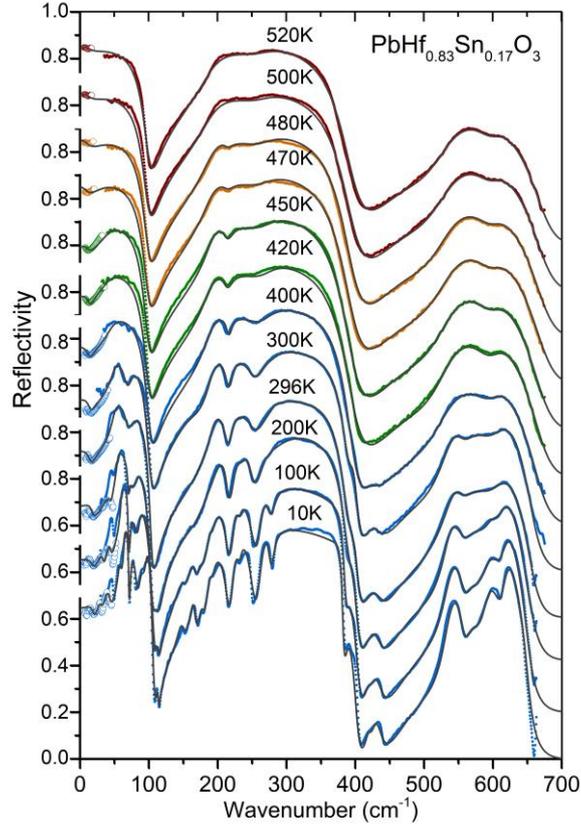

Figure 2: Far IR reflectivity spectra of PHS-17 at various temperatures on cooling (symbols), together with their fits (lines) using the eqs. (1,2). Uncertainty bars are smaller than the size of the symbols.

The simultaneous fit of the IR reflectivity and the complex dielectric spectra (ε' and ε'') from the THz transmission TDTTS measurements using the eqs. (1,2) helps to reveal additional excitations in the far-IR region, right below the phonon frequencies, which are difficult to detect relying only on the IR reflectivity spectra.

Both real ε' and imaginary ε'' parts of the dielectric permittivity, derived from the fits are displayed in Figure 3 at various temperatures, along with the experimental data obtained by THz spectroscopy.

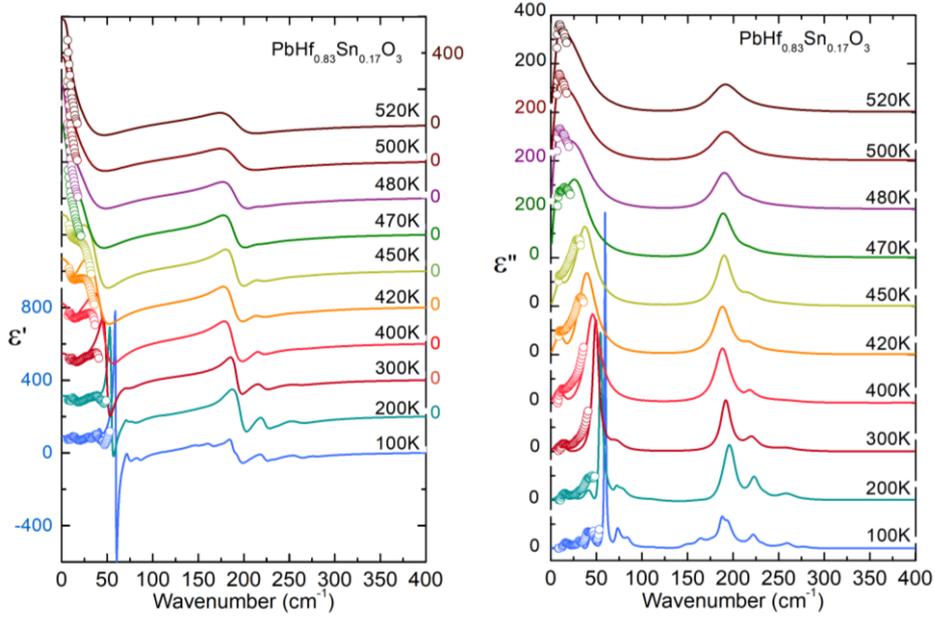

Figure 3: Real and imaginary parts of the complex permittivity of PHS-17 obtained from the fit (lines) together with experimental data from the THz transmission TDTTS experiment (symbols). Uncertainty bars are smaller than the size of the symbols.

From the PHS phase diagram [20], at 520 K the PHS-17 crystal should be in the cubic paraelectric phase, where there are three broad bands. As usual for lead-based complex perovskites, the bands show some substructure due to disorder. On cooling, the three main bands split into many phonons, indicating the change of symmetry at the phase transitions. New modes appear, in the reflectivity spectra, below 500 K, then below 400 K (see Figure 2). At the same time, there is a drop in the reflectivity values at low frequencies below 470 K. The imaginary part of the permittivity (Figure 3b) displays the shift of a peak at low frequencies (~30 cm$^{-1}$) towards higher frequencies near 60 cm$^{-1}$.

From the IR data and the analysis of the phonon parameters presented later in the discussion, two main transitions can be seen, below 470 K (from IM to AFE2) and below 400 K (from AFE2 to AFE1). The transition from cubic to the intermediate phase is not well seen at first sight.

### 3.3 Raman results

Raman scattering was measured in backscattering geometry from 800 K to room temperature to identify the non-polar phonons of PHS-17 in the different phases. The Raman spectra corrected from the Bose-Einstein population factor on the cooling cycle are depicted in Figure 4 a from 800 K to 300 K for parallel VV polarizers. They are shown in arbitrary counts, as the counts in the detector are not relevant anymore after the correction. The spectra in the cubic paraelectric phase are shown in red at the top of the figure, and the AFE1 phase is shown in blue in the lower part. Gradual changes are seen from 800 K down to ~460 K (red to yellow-green spectra) and then a clear new phase appear (green spectra). Below 400 K another abrupt change of the lines is seen, when entering into the AFE1 phase.

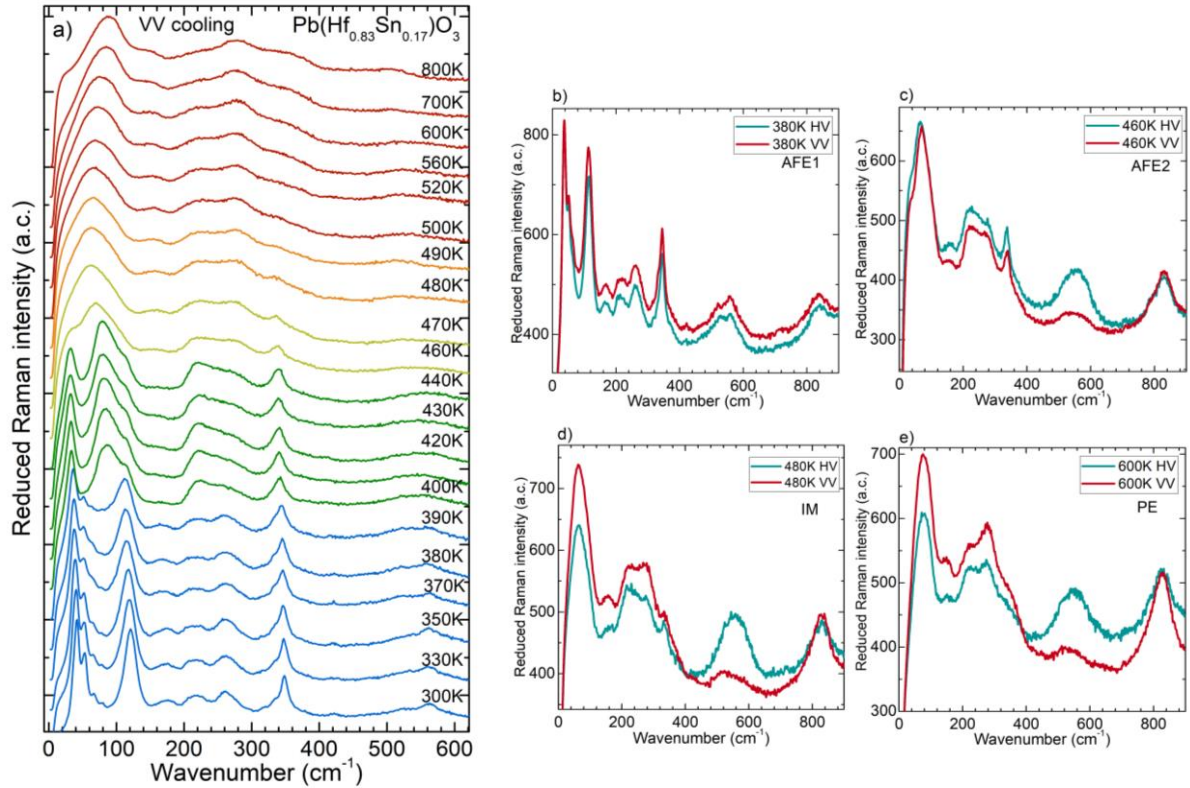

Figure 4: a) Raman spectra of PHS-17 with parallel polarizers. b) to e) Raman spectra of PHS-17 for parallel and crossed polarizers at selected temperatures in the four phases. All Raman spectra are corrected from the Bose-Einstein population factor, a.c. stands for arbitrary counts.

In panels b to e of Figure 4 the comparison of both geometries, VV and HV, is shown for selected temperatures. It is clear that the cubic phase is quite disordered and still shows some preferential orientation probably due to the formation of polar clusters within the cubic structure, because the Sn substitution strongly disrupts the cubic symmetry. The concomitant band near 830 cm$^{-1}$ at high temperatures resembles the $A_{1g}$ mode of the locally ordered $PbSc_{0.5}Nb_{0.5}O_3$ [30] and $PbSc_{0.5}Ta_{0.5}O_3$ [31], and it was related to the unit cell doubling in the cubic phase, transforming the $Pm\bar{3}m$ space group to the $Fm\bar{3}m$ one. This doubling implies ordering of the B atom at some correlation length detectable by Raman scattering. Therefore, PHS could also show traces of locally ordered Hf-Sn regions with this unit cell doubling.

The IM phase and the AFE2 phase show some anisotropy as well, although it is weaker in the AFE2 phase. And finally, the lowest AFE1 phase displays the same spectra in both cases, probably due to the random orientation of the domains in the polydomain crystal.

*3.4 Optical microscopy*

We performed optical imaging of the crystal under polarized light, and we were able to identify the phase transitions on heating and cooling. In PHS-17 the sequence of the phase transitions detected is the same as in PHS-30 [23]; although the IM phase has a narrower presence. We observed clear phase transitions and different developments of the phase transition sequence on heating and cooling. In addition, we were able to identify several types of domains developing within the sample, ferroelastic and antiferroelectric, with different dynamics. In Figure 5 we show several pictures taken during the cooling cycle at 5 K/min with parallel polarizers in a thin sample (thickness 0.38 mm). The sample is clearly transparent at 500 K in the cubic phase, and on cooling starts to develop the characteristic tweed pattern of the IM phase clearly seen at ~490 K. On further cooling through the IM phase the sample starts to develop very fine diagonal domains mixed with the IM phase (~470 K), probably ferroelastic domains, till it starts to change into the AFE2 phase (465.6 K) with very fine domain structure. This pattern evolves constantly on cooling till it settles. At 426.5 K the crystal shows the AFE2 phase, and at 404 K the AFE1 phase. The transitions on cooling showed two steps, with the ferroelastic domains developing faster and prior to the antiferroelectric ones. This might be related to the small anomalies seen near to the sharp ones in the calorimetry experiment (Figure 1a).

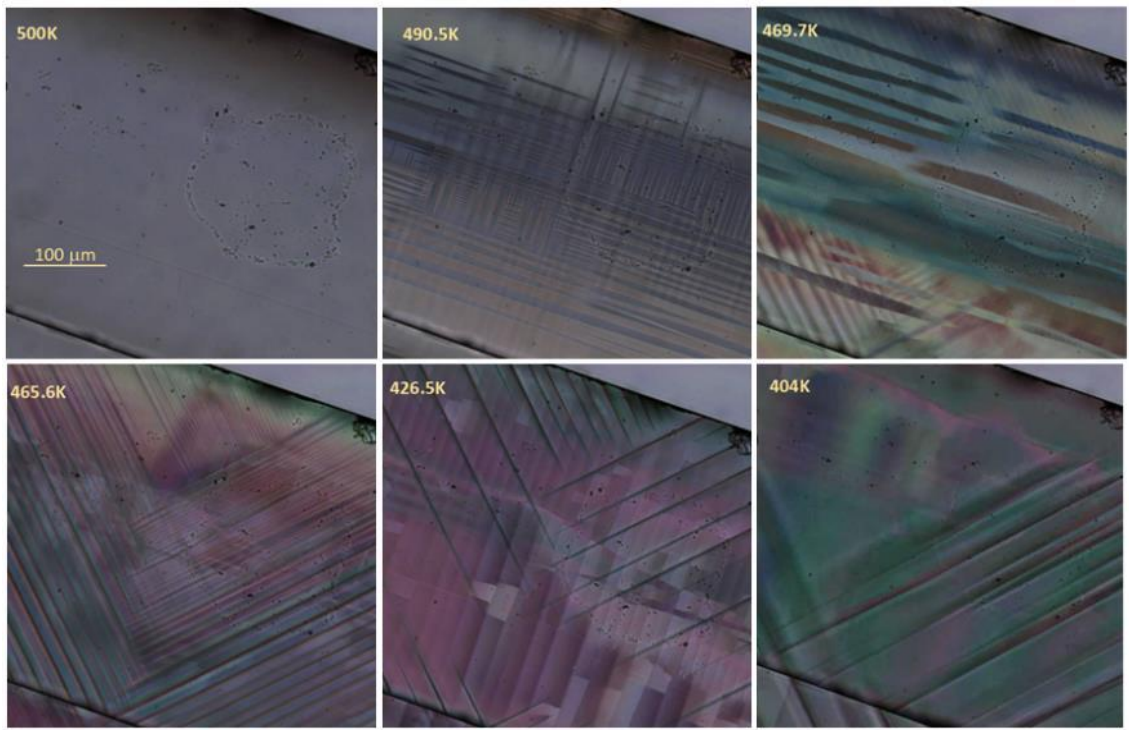

Figure 5: Optical images of the PHS-17 crystal illuminated in transmission with polarized light on cooling, at 500 K (cubic phase), at 490.5 K (IM phase), at 469.7 K (IM phase), at 465.6 K (AFE2 phase), at 426.5 K (AFE2 phase) and 404 K (AFE1 phase).

On heating, the transition between AFE2 and IM phases occurs even more clearly in two steps, and it shows a very fine parallel domain pattern with changing periodicity just before forming the tweed pattern of the proper IM phase (see Suppl. Material). In Figure 6 (Multimedia available online) we show the video of the transition sequence on the thicker crystal upon heating from 400 K (in the AFE1 phase) to 500 K (in the paraelectric phase).

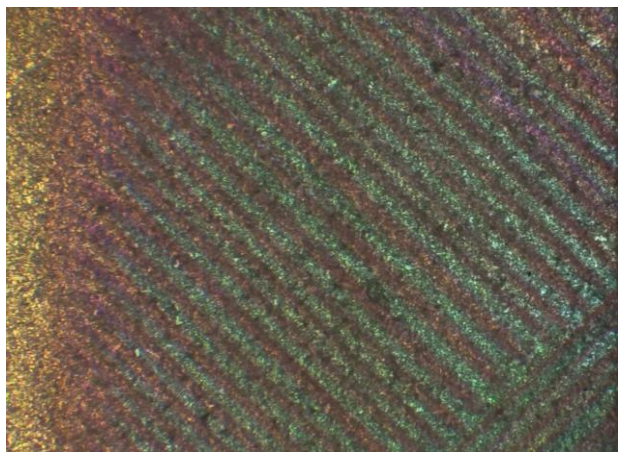

Figure 6: Frame at 474.5K of video imaging (Multimedia available online) of the thick PHS-17 crystal taken in transmission with polarized light on heating at a 10 K/min rate and using a 20x objective.

## 4. Discussion

All the experiments performed in PHS-17 crystals showed the sequence of phase transitions, and the temperatures found are summarized in Table 1. Some experiments revealed thermal hysteresis and dependence on the thermal rate. In Table 1 we present the approximate temperatures of the phase transitions found (heating/cooling format, except IR and THz experiments). DSC and dielectric experiments were performed at 1K/min, whilst IR, THZ and Raman spectroscopies plus optical microscopy were faster, at 10 K/min. We also found that the development of the phase transitions is more complex, showing two steps in the calorimetry curves. This effect is more evident on heating, but it could also be seen on cooling, mainly in the AFE1-AFE2 phase transition.

Table 1 Approximate temperatures (in Kelvin) of the phase transitions found in PHS-17 by various experiments on ↑heating/cooling↓. DSC and dielectric experiments performed at 1K/min. IR, THZ, Raman and optical microscopy at ~10 K/min.

|  | $T_C$ | $T_{IM-AFE2}$ | $T_{AFE2-AFE1}$ |
|---|---|---|---|
| DSC | ↑482/482↓ | ↑467/466↓ | ↑411/408↓ |
| Dielectric | ↑489/489↓ | ↑467/468↓ | ↑412/414↓ |
| Optical microscopy | ↑490/480↓ | ↑477/473↓ | ↑413/409↓ |
| Raman | 490-500 | 460-470 | 390-400 |
| IR | 490-500 | <470 | <420 |
| THz | -- | <470 | <420 |

*4.1 Symmetry considerations*

From the lattice dynamics point of view, it is more consistent to explain the phase transition sequence from the paraelectric PE phase, therefore we investigated mainly the cooling cycles. To study the transition from the cubic phase to the AFE2 phase, some simplifications need to be made because the AFE2 phase in PHS is incommensurate [14], [18] with averaged space group *Imma*. To include the disorder of the Pb atoms we considered the Pb atom in the cubic phase at the disordered site 8$g$ ($x,x,x$) [32], instead of the site (0,0,0). A symmetry analysis using SYMMODES tool [33] reveals the main lattice instabilities of the phase transitions of PHS. The irreducible representations for the lattice instabilities obtained are designated using the Miller-Love standard notation, where the numeric indexes mean different representations, involving different shifts of atoms and the indexes "+" or "−" indicate symmetry with respect to inversion symmetry at the origin.

Along the transition from the disordered cubic phase to the AFE2, there are two instabilities from the *R*-point (½, ½, ½) of the cubic Brillouin zone, labelled $R^{4+}$ and $R^{5+}$ related to the symmetric distortions of cations, with quadrupling of the unit cell [33]. As secondary instabilities, some modes at the *Γ*-point (0,0,0) could be activated too. The occurrence of a sequenced condensation of instabilities, as in PZT [34], could be the origin of the intermediate IM phase, prior to the phase transition to the AFE2 phase with *Imma* space group. A possible intermediate group could be the tetragonal *P*4/*mmm* driven by the *Γ*-point instability $Γ^{3+}$ (or GM3+), at the center of the Brillouin zone involving Pb shifts. (The orthorhombic *Cmmm* by condensation of another zone center mode $Γ^{5+}$ seems less probable as it has lower symmetry than *Imma*).

According to this analysis, we expect new soft modes in the AFE2 phase related to oxygen tilts and shifts of Pb atoms, but maybe also an extra mode in the IM phase related to a zone-center soft mode. The following transition from AFE2 to AFE1 *Pbam* is allowed mainly by the condensation of an instability from the Z-point (0,0,½) of the orthorhombic Brillouin zone. But, as the AFE2 phase is also incommensurate, there could be extra excitations from the incommensurate regions playing a role.

*4.2 Site group analysis*

The nature of the phase transitions can be better understood by analyzing the phonon parameters of the IR and Raman active phonons, which are different and complementary in centrosymmetric space groups.

The site-group analysis for PHS phases gives the following modes in the main phases, taking into account the corresponding multiplication of the unit cell.

PE ($Pm\bar{3}m$) Z=1: Γ = 4$F_{1u}$(**x,y,z**) + 1$F_{2u}$(-)

AFE2 (*Imma*) Z=4: Γ= 4$A_u$(-) + 10$B_{1u}$(**z**) + 7$B_{2u}$(**y**) + 9$B_{3u}$(**x**) + 7$A_g$(*xx,yy,zz*) + 7$B_{1g}$(*xy*) + 8$B_{2g}$(*xz*) + 8$B_{3g}$(*yz*)

AFE1 (*Pbam*) Z=8: Γ= 12$A_u$(-) + 12$B_{1u}$(**z**) + 18$B_{2u}$(**y**) + 18$B_{3u}$(**x**) + 16$A_g$(*xx,yy,zz*) + 16$B_{1g}$(*xy*) + 14$B_{2g}$(*xz*) + 14$B_{3g}$(*yz*)

Activities are in parenthesis: (-) for inactive, (**x**) or (**y**) or (**z**) for IR active, (*xx*), (*xy*), etc for Raman active phonons.

Due to the relatively high amount of Sn present, and due to the different masses of Hf and Sn, it is possible that the double occupancy of the B-site rises these numbers by the extra modes of Sn: 1$F_{1u}$(IR) in $Pm\bar{3}m$, (2$B_{1u}$+2$B_{2u}$+1$B_{3u}$) in *Imma*, and (3$B_{1u}$+3$B_{3u}$+3$A_g$+3$B_{1g}$+3$B_{2g}$+3$B_{3g}$) in *Pbam*. Therefore, Sn activates only IR new modes in the cubic and AFE2 phases (no new Raman modes), meanwhile in the AFE1 it could activate both.

In principle, for the averaged B-site we expect 3 IR active modes plus one acoustic in the cubic PE phase and no Raman active modes, 23 IR, 3 acoustic and 30 Raman modes in the AFE2 phase (for averaged *Imma*, Z=4) and 45 IR active, 3 acoustic and 60 Raman active modes in the AFE1 phase (*Pbam*, Z=8).

*4.3 Polar phonons*

Phonons detected by the IR experiments are polar, i.e. carry dipole moment and this splits the phonons into transverse optic (TO) and longitudinal optic (LO) components. However, in centrosymmetric materials, these phonons are not simultaneously Raman active, as seen from the mutually exclusive activities of each symmetry type. Therefore, in PHS both IR and Raman experiments measure different phonons and give complementary information about the crystal lattice throughout the phase transitions.

From the fit of the IR reflectivity spectra (Figure 2) the parameters of the polar phonons are obtained at the measured temperatures. The temperature dependence of the IR phonon frequencies is shown in Figure 7a, for TO and LO phonons.

In the cubic phase there are three main phonons, shown with larger symbols, three quite weak, related with disorder and locally broken cubic symmetry. In the low frequency part, an extra excitation, marked by stars, below the first phonon, was detected by THz measurements. There are at least two TO phonons showing a distinctive temperature dependence, related to the main Pb vibrations. The THz excitation is also displaying softening and hardening towards $T_{AFE2-IM}$. There are new phonons appearing on cooling; however, the new phases were fitted with less phonons than predicted by theory. The AFE2 phase was fitted with 8 and the AFE1 one with 27. This is quite common, as many of them are too weak to be detected or overlapped with others, mainly at temperatures above 400 K.

The dielectric strength of the phonons, depicted in Figure 7b, displays the contribution to the permittivity of the different excitations. The THz mode $\omega_{THz}$, below phonon frequencies, is a feature related with anharmonic Pb dynamics, and it is quite strong in the PE phase. On cooling its contribution diminishes and shows two anomalies near the transition temperatures. Therefore, it seems that this excitation is sensitive to all phase transitions and detects the different levels of disorder of Pb.

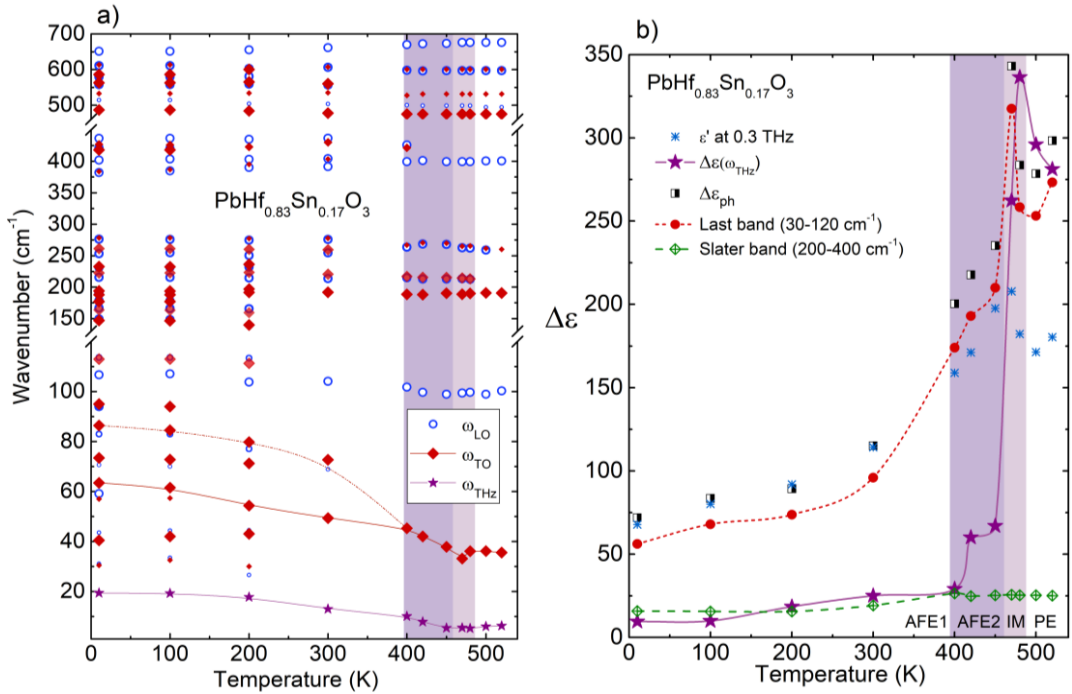

Figure 7: a) Temperature dependence of the frequencies of TO and LO phonons. b) Dielectric contribution to the permittivity of the THz excitation ($\omega_{THz}$), Last and Slater bands and the overall phonon contribution ($\Delta\varepsilon_{ph}$), together with experimental data at 0.3 THz. The two shaded regions in the panels highlight the two middles phases: AFE2 and IM.

The overall contribution of external phonons from the A-site (known as Last band) is also depicted and it shows important anomalies at $T_C$ and $T_{AFE2\text{-}IM}$, with a change in the slope in AFE1 phase. On the contrary, the contribution of the internal modes of the oxygen octahedra, Hf/Sn-O stretching vibrations is quite small, and it does not show anomalies related to the phase transitions. Nevertheless, there is a small change when entering the AFE1 phase, losing half of its contribution, from 26 to 15, which reveals that these modes are sensitive to the new Pb shifts in this phase. The values at 0.3 THz in the high temperature phases are lower than the contributions of the Last band and of the THz excitation, meaning that this frequency point is in between both driving mechanisms. On the contrary, in the lower AFE1 phase, the THz values correspond to the total value of phonons, which indicates that the THz excitation has lost its dielectric contribution and shifted to higher frequencies.

It is worthy to mention that dielectric measurements below THz frequencies were performed in another composition (with 8% of Sn) [35], where it was found that there is a contribution from an excitation softening in the 20–100 GHz range, which accounts for the values of the static permittivity in the PE and AFE2 phases. This frequency range has been related with domain wall vibrations in other ferroic materials [36]. This excitation is probably present as well in our composition but becomes negligible or vanishes in the AFE1 phase.

*4.4 Non polar phonons*

The non-polar phonons of PHS are detected by Raman scattering technique. This is important for this material, as the oxygen octahedra rotations and other modes out of the Brillouin zone are the main lattice instabilities producing the antiferroelectric phases. Already in the cubic phase we can see Raman signal —forbidden in the $Pm\overline{3}m$ space group— activated due to cationic disorder. The paraelectric PE phase of PHS was fitted using 9 bands and, on cooling, additional bands were necessary to fit the new modes appearing. The fit used damped harmonic oscillators and a quadratic baseline, which allows to fit more accurately some weak modes appearing on cooling which otherwise would be hidden in the background. The deconvolution of the fits, together with the experimental spectra for selected temperatures, is shown in Figure 8.

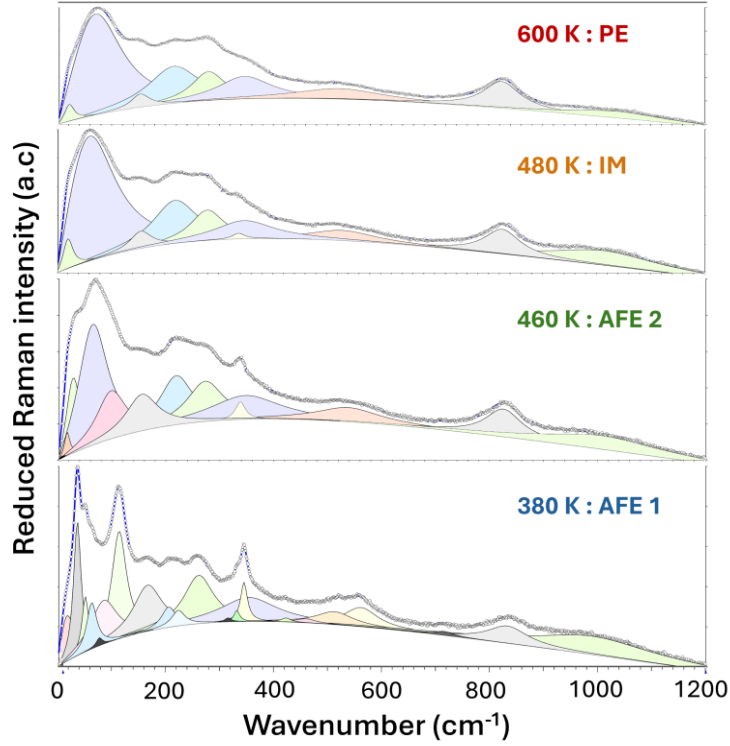

Figure 8: Deconvolution of the Raman fits (with damped harmonic oscillators and a quadratic baseline) together with the experimental spectra for PHS-17 at selected temperatures in the four phases. Raman spectra are corrected from the Bose-Einstein population factor, a.c. stands for arbitrary counts.

The temperature dependences of the non-polar mode frequencies are presented in Figure 9 together with the damping constants and areas of the main bands.

Five main modes (and four weaker ones) are present at high temperatures, even in the cubic phase, as detected in other compositions. Two modes at lower frequencies (labelled Pb1 and Pb2) originate from the Last band, demonstrating that Pb atoms are highly disordered and fluctuate at a local level, as observed in $PbHfO_3$ [31]. On cooling these modes show an important softening towards $T_C$. The higher frequency one (~70 cm$^{-1}$, Pb2) is overdamped and we show in Figure 9a its renormalized frequency $\omega_r = \omega^2/\gamma$, where $\gamma$ is the damping constant.

Below $T_C$ a new mode emerges at 350 cm$^{-1}$, becoming more intense on cooling to the AFE2 phase. This might indicate the coexistence of the IM and AFE2 phases during the nucleation of the AFE2 phase.

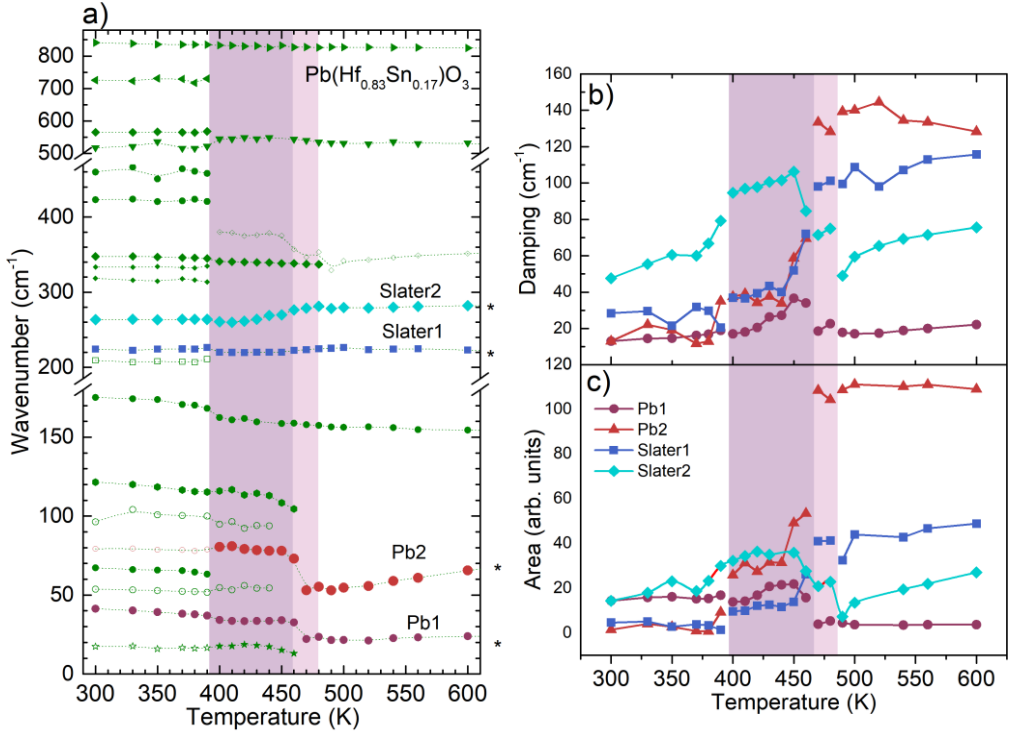

Figure 9: a) Frequencies of the Raman phonons of PHS-17. b) Damping constants and c) areas of the four main bands. The two shaded regions in the panels highlight the two middles phases: AFE2 and IM.

Below $T_{\text{IM-AFE2}}$ there is a new mode that appears at ~120 cm$^{-1}$, which is assigned to the antiferrodistortive soft mode from the R point. This mode is related to the antiphase rotation of oxygen octahedra in adjacent unit cells, as observed in PbZrO$_3$ [5]. This mode becomes stronger on cooling and remains in the AFE1 phase. The modes labelled as Pb1 and Pb2 here correspond to the modes *b* and *d* of previous studies [23]. The antiferrodistortive mode corresponds to the one labelled as *e*.

Another extra low-frequency mode is found below $T_{\text{IM-AFE2}}$, marked by stars in Figure 9a. This mode appears as a shoulder in the lowest frequency Raman band. Its frequency is not accurate enough to obtain a physically convincing temperature dependence, but it may be related to the antipolar shifts of Pb atoms.

A broad feature, present in the spectrum at around 150 cm$^{-1}$ at both low and high temperatures, is probably related to disorder caused by fluctuations of both oxygen tilts and Pb atoms [19]. The other two main bands in the 200–300 cm$^{-1}$ range originate from the Slater perovskite band, resulting from vibrations of the B-site interacting with the oxygen octahedra.

In panels b and c of Figure 9, the respective damping constants and areas of these main bands show anomalies in cooling, reflecting the changes of phases. The presence of the intermediate IM phase is more noticed by the two Slater bands, displaying jumps in the areas and damping constants at $T_\text{C}$, especially seen in the mode with frequency near 280 cm$^{-1}$. On the contrary, Pb related vibrations are more sensitive to the transition to the AFE2 phase, this is better seen in the abrupt change of area and damping constant below $T_{IM-AFE2}$. These two vibrations experience further changes upon cooling into the AFE1 phase. The mode labelled Pb2 loses almost its whole area and becomes hidden and practically invisible in the spectra. This was seen also in other PHS compositions [21]. The higher sensitivity of the Slater bands to the IM phase can be related to the fact that this phase transition is mainly triggered by oxygen tilts, whereas the Pb atoms are still too dynamically disordered to respond coherently to it. Pb atoms probably exhibit both polar and antipolar fluctuations in the IM phase. However, when the antipolar fluctuations prevail, their vibrations can propagate resulting in a significant change in their phonon frequencies, as observed in the Raman spectra.

*4.5 Comparison of polar and nonpolar phonons*

To compare the behaviour of polar and nonpolar phonons, we plot (Figure 10) together the Pb related vibrations (below 120 cm$^{-1}$), the antiferrodistortive mode, and the Hf/Sn-O vibrations from the Slater band at middle frequencies (200–400 cm$^{-1}$). Both types of vibrations A-site related (Last, Figures 10a,b) or B-site related (Slater, Figures 10c,d) "perceive" the phase transitions in a different way.

The main Pb vibration seen by IR spectroscopy shows a drop in its frequency and a maximum of its contribution to permittivity or dielectric strength in the IM phase (at 460 K). This points to a change in the Pb dynamics when entering this phase. A sudden drop in its dielectric strength happens when the crystals enter into the AFE2 phase. These two effects mean that Pb vibrations are highly anharmonic in the IM phase contributing more to the permittivity and then at $T_{IM-AFE2}$ the atoms begin to order into an antipolar fashion. The THz mode displays a maximum already at 470 K. So, the IM phase is detected earlier at lower frequencies, where the extra anharmonic dynamics are present.

The Pb vibrations detected by Raman, on the other hand, soften markedly with a small anomaly in the IM phase and then a jump after entering the AFE2 phase, when a new low frequency mode appears together with the antiferrodistortive mode, labelled AFD in the figure. The integrated area of the Raman spectra (normalized to the background) below 140 cm$^{-1}$ shows a maximum in the AFE2 phase mainly due to the Pb2 vibration.

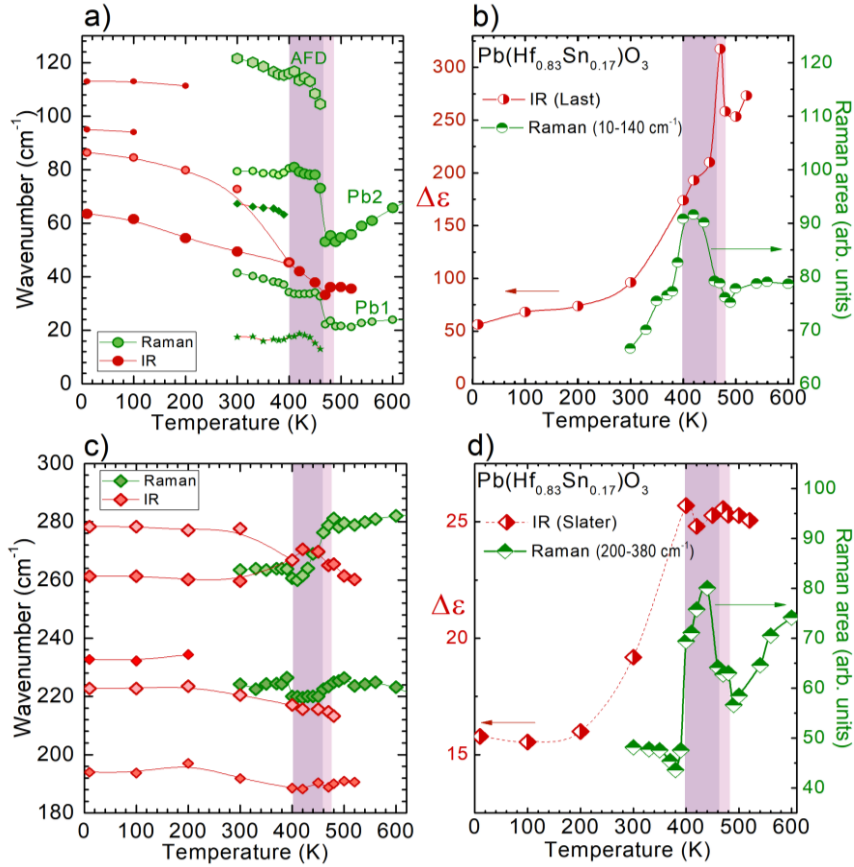

Figure 10: a,c) Frequencies of the IR (in red) and Raman (in green) phonons, and b,d) dielectric strength and Raman integrated area, for the Last band below 140 cm$^{-1}$ (Pb vibrations, upper panels, labelled as Pb1 and Pb2) and for the Slater band (200–400 cm$^{-1}$, lower panels). AFD stands for the antiferrodistortive soft mode. The two shaded regions in the panels highlight the two middles phases: AFE2 and IM.

The middle frequency modes from the Slater band detected by IR and Raman spectroscopies behave differently. The polar modes in this region are practically not sensitive to the IM and AFE2 phases, and their combined dielectric strength shows a monotonous trend, till it drops at $T_{AFE2-AFE1}$. This proves that the two transitions at $T_C$ and $T_{IM-AFE2}$ are not related to polar modes involving B-site cations. On the contrary, the Raman modes detect both transitions, proving that the driving instabilities are non-polar and related to oxygen vibrations. The integrated area of the Slater band shows two consecutive jumps on cooling, at $T_C$ and $T_{IM-AFE2}$, and then slowly decreases, with a steep drop in the AFE1 phase, when the unit cell doubles again.

From the combined IR, THz and Raman spectroscopic investigations, the sequence of phase transitions in PHS-17 has been related to the dynamics of lattice vibrations of different origins.

The parameters of the lowest-frequency phonons suggest that Pb atoms are sensitive to all transitions and their dynamics change with temperature. At high temperatures, the paraelectric phase exhibits heavily anharmonic fluctuations. However, on cooling the fluctuations begin to organize, first forming polar and antipolar fluctuations (producing the IM phase), and finally producing antipolar shifts (in the AFE2 phase).

The intermediate IM phase acts as a transitional phase that develops in order to accommodate the disorder of Pb atoms and the fluctuations of oxygen tilts prior to the formation of the AFE2 phase. From lattice dynamics arguments, this could be partially caused by soft tilts from the R-point of the cubic Brillouin zone ($R^{4+}$ and $R^{5+}$), together with another instability from the Brillouin zone center ($\Gamma^{3+}$) allowing tetragonal $P4/mmm$ symmetry. The polar and antipolar Pb fluctuations in the PE and IM phases, mixed with the fluctuation of tilts [18], finally transform into antipolar shifts (incommensurate at first, and then stabilized), probably under the influence of the $R^{4+}$ and $R^{5+}$ modes, which take the role of order parameters, quadrupling the unit cell. The transition PE-IM exhibits a combination of displacive and order-disorder characters. The intermediate IM phase also shows traces of coexistence of the paraelectric and AFE2 phases.

On further cooling, the condensation of a lattice instability from the Z-point (0,0,½) of the orthorhombic Brillouin zone drives the PHS crystal into the AFE1 phase with $Pbam$ space group with another doubling of the unit cell. This phase transition AFE2-AFE1 is prevailingly displacive and of first-order character, as demonstrated by the presence of thermal hysteresis.

The influence of Sn atoms on the phase transition is seen only qualitatively in the presence of the IM phase. To understand properly the atomistic mechanism behind it we would need to measure all phonons in oriented single crystals of PbHfO$_3$ of a reasonable size, which is still a pending task, and compare phonons in several PHS crystals with different Sn concentrations.

For PHS-17, the Sn concentration could be not enough to see important differences in phonons or detect the double occupancy of the B-site. However, we have seen that a new Raman mode near 350 cm$^{-1}$ is detected in the intermediate IM phase. And the temperature behavior of the THz excitation shows some anomalies related to this intermediate phase.

At high temperatures, the band near 830 cm$^{-1}$ could be related to the unit cell doubling in the cubic phase due to the presence of Sn in the B site and the change from $Pm\bar{3}m$ to the $Fm\bar{3}m$ space group. This doubling implies ordering of the B atom at some correlation length detectable by Raman scattering. The atomistic origin of the Raman bands at these frequencies in perovskites is traditionally linked to the breathing modes of the oxygen octahedra, and some non Pb-based perovskites display these modes even without B-site ordering [37]. Therefore, more detailed atomistic calculations of the Raman spectra in PHS are needed in order to shed light on the quantitative influence of Sn on the phase transition sequence.

# 5. Conclusions

The complementary phonon studies of PHS-17 by means of THz, far infrared, and Raman spectroscopies revealed that the phase transitions are of phononic origin due to the condensation of several lattice instabilities. The different selection rules for IR and Raman phonons allow to detect polar and non-polar vibrations, to fully characterize the phase transitions.

The overall transformation from the cubic paraelectric phase to the antiferroelectric orthorhombic *Pbam* phase enables two middle phases in order to accommodate sequentially the shifts of Pb atoms and tilts of the oxygen octahedra.

The paraelectric PE and intermediate IM phases are distorted from the cubic symmetry exhibiting different levels of disorder. In the PE phase, random fluctuations of Pb atoms give rise to anisotropic Raman scattering, and IR spectra show a weak substructure in the main reflectivity bands. Upon cooling, the IM phase (characterized by an intertwined domain pattern) is detected by a THz anharmonic excitation related to the fluctuation of atoms, by a polar and an antipolar Pb vibration, and by non-polar Hf/Sn-O vibrations and oxygen octahedra tilts. Symmetry considerations suggest that this phase has an averaged tetragonal symmetry but exhibits local polar distortions.

On further cooling the Pb fluctuations transform gradually into antipolar shifts, whose vibrations are detected by IR and Raman spectroscopy in the AFE2 phase. The antiferrodistortive modes and the B-site related modes in the 200–400 $cm^{-1}$ region primarily drive this phase transition. The same happens at the last transition to the AFE1 phase, characterized by sudden jumps in the frequencies of the antiferrodistortive and antipolar soft modes.

The Pb atom is more sensitive to the high temperature phase transitions, since its dynamics strongly change with temperature causing the main dielectric anomaly. Oxygen atoms play a more significant role in the AFE2 and AFE1 phases, because the antiferrodistortive soft modes are driving these transitions.

Optically, the domain dynamics suggest that the presence of Sn triggers the development of the IM phase with its distinctive tweed pattern. Ferroelastic and antiferroelectric domains coexist, but they evolve differently with temperature, causing two-steps phase transitions. In the AFE2 phase, the ferroelastic domains seem to be affected by the incommensurate modulations. Only when these modulations stabilize do the antiferroelectric domains fully develop in the crystal.

**Supplementary Material**

We performed optical imaging of a thick PHS-17 crystal under crossed polarized light on heating at the rate of 10 K/min, and results of this experiment are shown in Supplemental material number 1 (description of the experiment and pictures).

**Acknowledgments**

This work was partially financed (A.K.R. and E.B) by the Czech Science Foundation (GAČR) through a Lead Agency bilateral Czech-Slovenian project (no. 24-10699K).

**Data availability statement**

The data that support the findings of this study are available from the authors upon reasonable request.

**Disclosure statement**

No potential conflict of interest was reported by the authors.